\documentclass[prl,preprint,endfloats]{revtex4-2}

\usepackage{graphicx}
\usepackage{bm}
\usepackage{pifont}
\usepackage{amsmath}
\usepackage{color}
\usepackage{comment}
\usepackage{textcomp}

\begin{document}

\title{Piezomagnetic transport in van der Waals noncoplanar Antiferromagnets}

\author{Abdul Ahad$^{1,a}$, Miuko Tanaka$^{1}$, Nguyen Duy Khanh$^{2}$, Riku Ishioka$^{2}$, Aki Kitaori$^{2}$, Tenta Kitamura$^{3}$, Hao Ou$^{3}$, Jiang Pu$^{3}$, Shinichiro Seki$^{2}$, Toshiya Ideue$^{1,a}$} 

\affiliation{$^{1}$Institute for Solid State Physics, The University of Tokyo, 5-1-5 Kashiwanoha, Kashiwa, Chiba, Japan, 
\footnotetext[1]{Corresponding authors.
		abdulahad@issp.u-tokyo.ac.jp,
		ideue@issp.u-tokyo.ac.jp
	}
	$^{2}$Department of Applied Physics and Quantum-Phase Electronics Center, The University of Tokyo, Tokyo, Japan,\\
	$^{3}$Department of Physics, Institute of Science Tokyo, Tokyo, Japan}

\begin{abstract}
{\bf The piezomagnetic effect—strain-induced linear modulation of magnetization, arises in magnets with broken time-reversal symmetry (BTRS), offering a pathway to bidirectional strain-based control of magnetism, which is an essential straintronic and spintronic functionality in solids. Metallic antiferromagnets with BTRS provide an ideal platform to study this effect through transport measurements, yet experimental demonstrations are limited. Van der Waals (vdW) nanomagnets, with their mechanical flexibility, are particularly promising for realizing large piezomagnetic responses and effective transport control.
	Here we demonstrate piezomagnetic control of electronic transport in nano-devices of the vdW antiferromagnets CoNb$_3$S$_6$ and CoTa$_3$S$_6$, archetypal vdW metals with BTRS that exhibit a spontaneous Hall effect. Applying uniaxial strain linearly modulates both the antiferromagnetic transition temperature and coercive field, consistent with strain-driven tuning of exchange coupling, key signatures of the piezomagnetic effect. Moreover, spontaneous Hall effect is controllable via strain, evidencing piezomagnetic tuning of Berry curvature and its associated geometric transport. These findings establish piezomagnetism as a powerful route to manipulate antiferromagnetic transport, opening avenues for straintronic and spintronic applications in vdW magnetic systems.
	
}
\end{abstract}

\maketitle
\clearpage
\section{Introduction}
Antiferromagnetic materials provide unique advantages over conventional ferromagnets, including ultrafast spin dynamics and negligible stray fields. Yet, the ability to control their magnetic states remains a central challenge for harnessing spin degrees of freedom in solids. In particular, antiferromagnets with characteristic symmetry breaking such as broken time-reversal or inversion symmetry, enable controlling magnetic properties by external magnetic or electric fields, and have therefore attracted growing attention in recent years~\cite{Wadley2016,Tsai2020,KhaliliAmiri2024,Liu2019,Baltz2018,Jungwirth2016,Takeuchi2025}.
Strain offers another powerful means of magnetic control. The piezomagnetic effect, first proposed by Voigt~\cite{Voigt1928} and later formalized by Dzyaloshinsky~\cite{Dzialoshinskii1954}, describes a linear response of magnetization ($M$) to applied strain ($\epsilon$), $i.e.$ $M \propto \epsilon$. This effect is allowed in magnets lacking time-reversal symmetry~\cite{Borovik-Romanov1994}, and has been observed in insulating antiferromagnets such as collinear~\cite{Borovik-Romanov1960} CoF$_2$ and MnF$_2$—now recognized as altermagnets~\cite{Cheong2025}, and in noncollinear systems~\cite{Baryakhtar1985,Jaime2017} such as UO$_2$. In metallic antiferromagnets, piezomagnetic modulation of electronic states can influence transport properties such as the Berry-curvature-induced anomalous Hall effect, which can be detected even in nanodevices where direct magnetization measurements are challenging. Only a few antiferromagnetic metals with BTRS, such as Mn$_3$NiN, Mn$_3$Ir, and Mn$_3$Sn, have been reported to exhibit such strain-tunable transport~\cite{Boldrin2018,Zhang2025,Ikhlas2022}, but previous studies were restricted to small strain levels ($\leq 0.3\%$), leaving the microscopic mechanisms and full potential of piezomagnetism largely unexplored.
Van der Waals (vdW) antiferromagnets offer a transformative platform to overcome these limitations. Their mechanical flexibility enables large elastic strains of several percent~\cite{Dong2023}, while their reduced dimensionality and characteristic spin textures give rise to rich geometric transport phenomena~\cite{Deng2020}. Here we establish vdW noncoplanar antiferromagnets~\cite{Ikhlas2022} as a model system for exploring and exploiting piezomagnetic control of transport. Using CoNb$_3$S$_6$ and CoTa$_3$S$_6$ nano-devices, we demonstrate that uniaxial strain linearly modulates both the antiferromagnetic transition temperature and coercive field, consistent with strain-driven tuning of exchange interactions. Crucially, we show that the spontaneous Hall effect can be tuned by strain, providing direct evidence of piezomagnetic control of Berry curvature and geometric transport. These results indicate piezomagnetism as a powerful principle for manipulating exotic transport properties in metallic antiferromagnets and highlight vdW magnets as a versatile platform for straintronic and spintronic applications~\cite{Dong2023,Liu2024,Ou2025}.

\section{Results} 

\section{Antiferromagnetic state with broken time reversal symmetry and Spontaneous Hall effect in \NoCaseChange{CoNb$_3$S$_6$}}
\begin{figure}
\begin{center}
\includegraphics*[width=13cm]{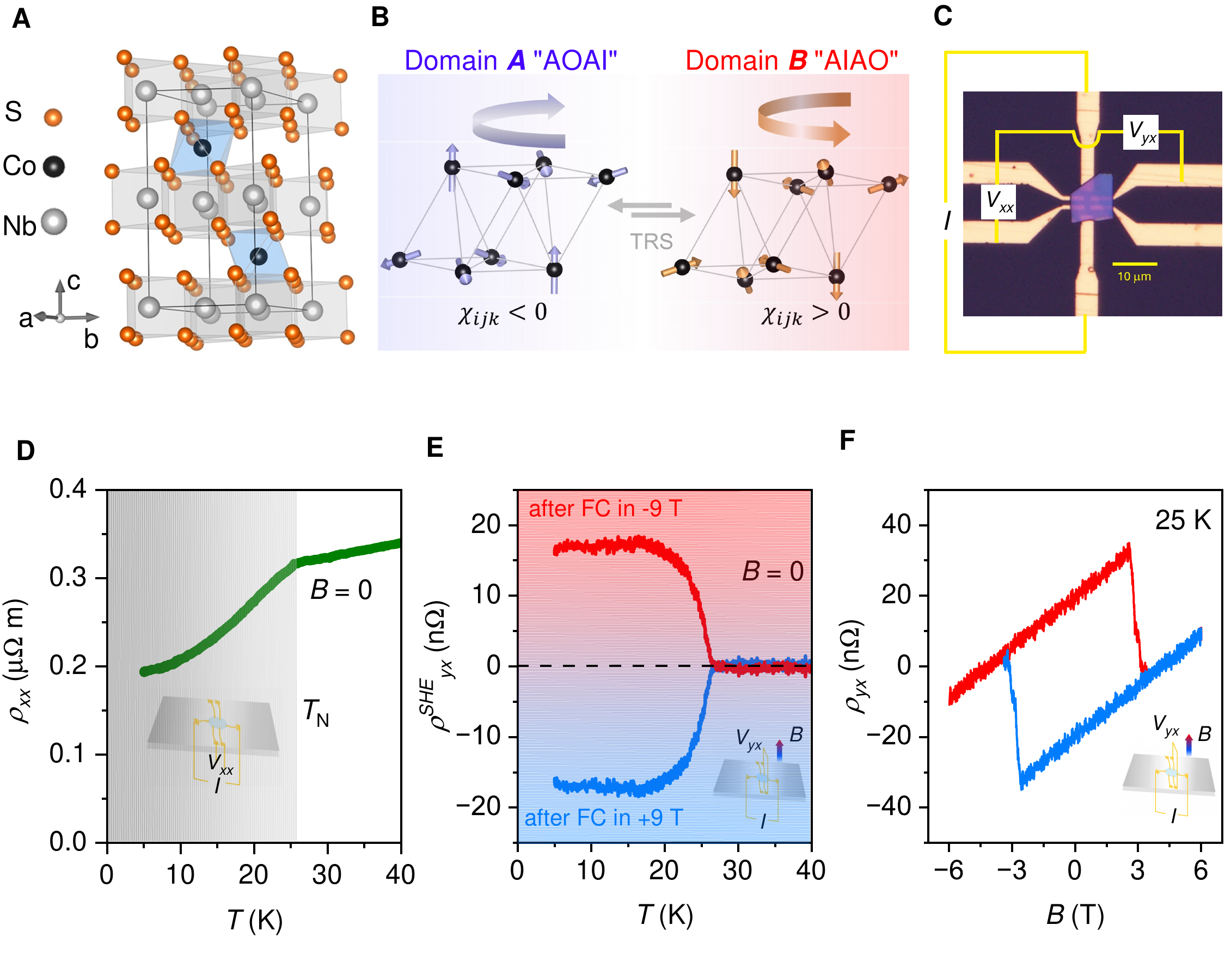}
\caption{{\bf Crystal structure, magnetic order, and the spontaneous Hall effect in van der Waals antiferromagnet CoNb$_3$S$_6$}\\ {\bf (A)} Schematic of the crystal structure of CoNb$_3$S$_6$. {\bf (B)} Two distinct non-coplanar antiferromagnetic orders with broken time reversal symmetry (A and B domains). Each domain shows opposite sign of scalar spin chirality ($\chi_{ijk}$) and resultant spontaneous Hall effect. {\bf (C)} Optical microscope image of a nano-device fabricated on flexible substrate for electronic transport measurements. {\bf (D)} Temperature dependence of zero field ($B$ = 0) longitudinal resistivity ($\rho_{xx}$). Kink corresponds to the antiferromagnetic transition at Neel temperature, $T_N$. {\bf (E)} Temperature dependence of spontaneous Hall resistivity ($\rho_{yx}^{SHE}$) under $B$ = 0. Red and blue curves represent the different domain contributions selected by field cooling ($B \parallel c$) process. {\bf (F)} Magnetic field dependence of $\rho_{yx}$, recorded just below $T_N$ (at 25 K). The red and blue curves correspond to magnetic-field switchable domains via out of plane magnetic field. Insets of (D), (E), (F): illustrate the experimental configurations.}
\label{Fig1}
\end{center}
\end{figure}

Figure~\ref{Fig1}A represents the crystal structure of CoNb$_3$S$_6$. It consists of NbS$_2$ triangular layers with Co ions intercalated between the van der Waals gaps of triangular layers, resulting in non-centrosymmetric chiral structure with space group $P$6$_3$22~\cite{Khanh2025,Takagi2023,Park2023,Ghimire2018}. It shows non-coplanar antiferromagnetic spin ordering below $T_N$ $\sim$ 26 K in which Co atoms forms All-in-All-out (AIAO) spin structure characterized by magnetic space group $P$32'1. In this AIAO spin configuration, the Co moments cancel each other, providing almost vanishing total magnetization per unit cell $i.e.$, $M_s=\sum_{i=1}^{4} m_i \approx 0$. Despite its negligible spontaneous magnetization, the AIAO spin configuration possess finite scalar spin chirality $\chi_{ijk}=S_i.(S_j \times S_k)$ that breaks time reversal symmetry, allowing non-zero spontaneous Hall effect~\cite{Takagi2023,Park2023}. The other time reversal pair of AIAO spin configuration have opposite sign of $\chi_{ijk}$ and hence opposite Berry curvature-induced spontaneous Hall effect (see Fig.~\ref{Fig1}B).  We transferred the exfoliated CoNb$_3$S$_6$ on flexible substrate with bottom electrodes and studied the transport properties. Figure~\ref{Fig1}D and E shows the temperature dependent longitudinal ($\rho_{xx}$) and spontaneous transverse resistivity ($\rho_{yx}$) of Device $\#$1 ($t$ = 91 nm, Fig.~\ref{Fig1} C). $\rho_{xx}$ under zero magnetic field captures the onset of $T_N$ while $\rho_{yx}$ ($B$ = 0) exhibit spontaneous Hall effect whose sign depends on the different antiferromagnetic domains related by time reversal symmetry (domains A and B). This domain selection is possible by cooling the sample in +9 T and -9 T however data was recorded in zero field while increasing the sample temperature.  Figure 1F shows the magnetic field dependence of $\rho_{yx}$ measured under $B \parallel c$ at 25 K.  Ferromagnetic-like hysteresis with coercive field around 3 T represents that antiferromagnetic A and B domains can be switchable by applying the magnetic field. The additional positive slope comes from the ordinary Hall effect, $i.e.$, $\rho_{yx}=R_o B + \rho_{yx}^{SHE}$, where $R_o$ is the ordinary Hall coefficient, $B$ is magnetic field, and $\rho_{yx}^{SHE}$ is the spontaneous Hall resistivity. The sign of $R_o$ indicates the holes as majority charge carrier and its magnitude is unchanged by the temperature variation~\cite{Ghimire2018}. These behaviors are consistent with the previous bulk study~\cite{Takagi2023}.

\section{Strain modulation of antiferromagnetic ordering}
\begin{figure}
\begin{center}
\includegraphics*[width=15.0cm]{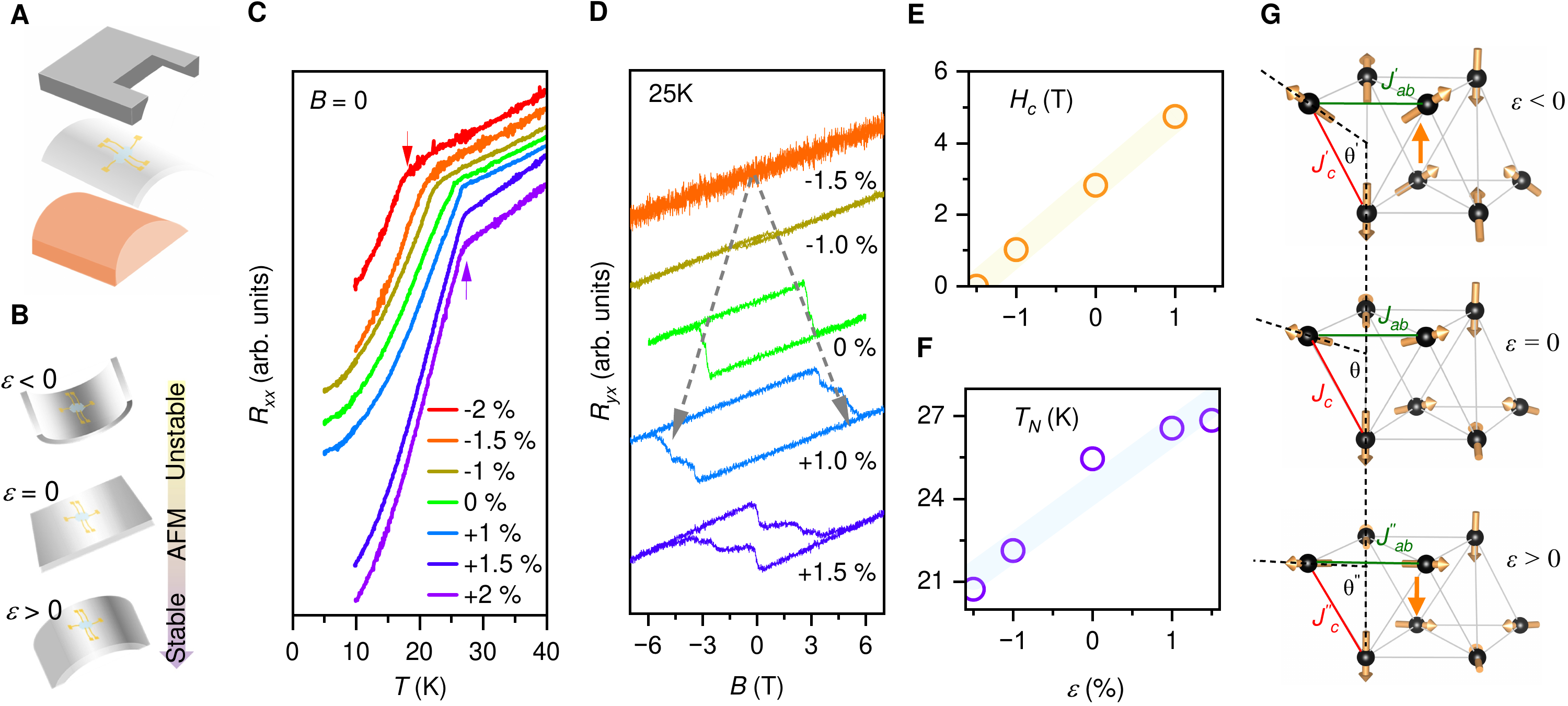}
\caption{{\bf Strain modulation of antiferromagnetic state in CoNb$_3$S$_6$.} \\
{\bf (A)} Experimental set-up for strain application using bending stage and flexible substrate. Flexible polyethylene naphthalate (PEN) substrate, which contains the exfoliated flakes of CoNb$_3$S$_6$, is sandwiched between a curved Cu-stage and a fixture. {\bf (B)} Schematics of the application of uniaxial strain. By changing the radius of curvature of substrate, we can achieve both tensile ($\epsilon$ $>$ 0) and compressive strain ($\epsilon$ $<$ 0), as well as unstrained state ($\epsilon$ = 0). {\bf (C)} Temperature dependence of longitudinal resistance under various applied uniaxial strain in $B$ = 0. Red and purple arrows indicate $T_N$ for -2 \% and +2 \% strains, respectively. {\bf (D)} Magnetic field dependence of $R_{yx}$ under applied strain at 25 K. Dashed grey arrow is a guide to the eye to show Hc evolution with strain magnitudes. {\bf (E)}, {\bf (F)} Strain dependence of the $H_c$ (at 25 K) and $T_N$ (at $B$ = 0), respectively. {\bf (G)} Schematics of spin configurations and their modulations under strain. Top, middle and bottom figures correspond to compressive ($\epsilon$ $>$ 0), unstrained ($\epsilon$ = 0) and tensile ($\epsilon$ $<$ 0) strain, respectively. The canting angle, and exchange interactions are assumed to be modulated under strain as in the previously reported case of piezomagnetic effect in non-collinear spin structures (see supplementary materials note 7). 
}
\label{Fig2}
\end{center}
\end{figure}

Aforementioned antiferromagnetic states with BTRS in CoNb$_3$S$_6$ allows the piezomagnetic control of transport properties. We applied uniaxial strain on CoNb$_3$S$_6$ flake samples by using the setup illustrated in Fig.~\ref{Fig2}A.  The exfoliated samples of CoNb$_3$S$_6$ are mounted on flexible substrates and fixed on the bending stage with various radius of curvature. (see Materials and Methods for detail and supplementary note 1). Both types of uniaxial strain i.e., compressive ($\epsilon$ $<$ 0) and tensile ($\epsilon$ $>$ 0) can be applied by using this method (see Fig.~\ref{Fig2} B). Figure~\ref{Fig2}C and D demonstrate the temperature dependence of the resistance $R_{xx}$ (T) under zero magnetic field and magnetic field dependence of Hall resistance $R_{yx}$ (B) at 25 K under several fixed strains (-0.02 $\leq \epsilon \leq$+0.02). Both show the systematic modulation of the antiferromagnetic phase under uniaxial strain, demonstrating the trend that the tensile strain favors the antiferromagnetic state. Especially, it is noted that huge change in  $H_c$ with the strain may be useful for straintronic memory applications~\cite{Dieny2017} (see supplementary note 11). Plotting the antiferromagnetic phase transition temperature $T_N$ and coercive field $H_c$ as a function of strain value reveals that they are linearly modulated with strain (Figs.~\ref{Fig2}E and F). 
This linear change of $T_N$ and $H_c$ can be related with piezomagnetic effect through the strain-induced modulation of the exchange interaction as follows. On the one hand, the magnetic point group of CoNb$_3$S$_6$ (32') allows the generation of out-of-plane magnetization ($M_z$) in response to in-plane strain ($\epsilon_{ij}$), $i.e.$, $M_z= \Lambda_{zxx} \epsilon_{xx}+\Lambda_{zyy} \epsilon_{yy}+\epsilon_{zzz} \epsilon_{zz}$, where $\Lambda_{ijk}$ is a piezomagnetic tensor~\cite{Gallego2019}. The linear coupling between magnetization and the strain microscopically originates from the strain-induced modulation of the exchange interaction $J_{ij}$ ($\epsilon$), $i.e.$, $\Delta J$. In case of antiferromagnets with triangular lattice described by Heisenberg model, $\Delta J$ is proportional to the strain, resulting in strain-induced linear magnetization~\cite{Zhang2025,Lukashev2008,Zemen2017,Boldrin2018}. In the present case, strain-induced imbalance of  $J_{ij}$  causes the canting of spin angular momentum thereby eliminating complete cancellation and generating magnetization (Fig.~\ref{Fig2}G, see also supplementary material note 7). It should be noted here that the results do not seem to depend particularly on the in-plane direction of the uniaxial strain.  (see supplementary material note 3). This observation is in excellent agreement with the same piezomagnetic tensor for $\epsilon_{xx}$ and $\epsilon_{yy}$, i.e., $\Lambda_{zxx}$ = $\Lambda_{zyy}$ (see supplementary note 6 for details).
On the other hand, this strain-induced modulation of the exchange interaction $\Delta J$ can also explain the linear dependency of $T_N$ and $H_c$ on strain. For example, according to the mean field approximation, transition temperature is expressed as  $T \propto J⁄k_B $, where $k_B$ is the Boltzmann constant, naturally leading to the linear modulation of $T_N$.  The change of $H_c$, which is also observed in other frustrated magnetic systems, can be attributed to the strain-tunable magnetic anisotropy with strain~\cite{Ikhlas2022}. Implicitly, the magnetic anisotropy energy ($E_A$) is relevant to the piezomagnetic effect as $\Lambda_{ijk}=\sum_j \frac{\partial^2 E_A}{\partial \sigma_{jk}\,\partial B_i}$~\cite{Phillips1967}. Strain-modulated crystal symmetry and local magnetic moment rotations result in alterations in magnetic anisotropy, eventually allowing effective modulation of the coercivity.

\section{Piezomagnetic control of Berry curvature and spontaneous Hall effect}
\begin{figure}
\begin{center}
\includegraphics*[width=15cm]{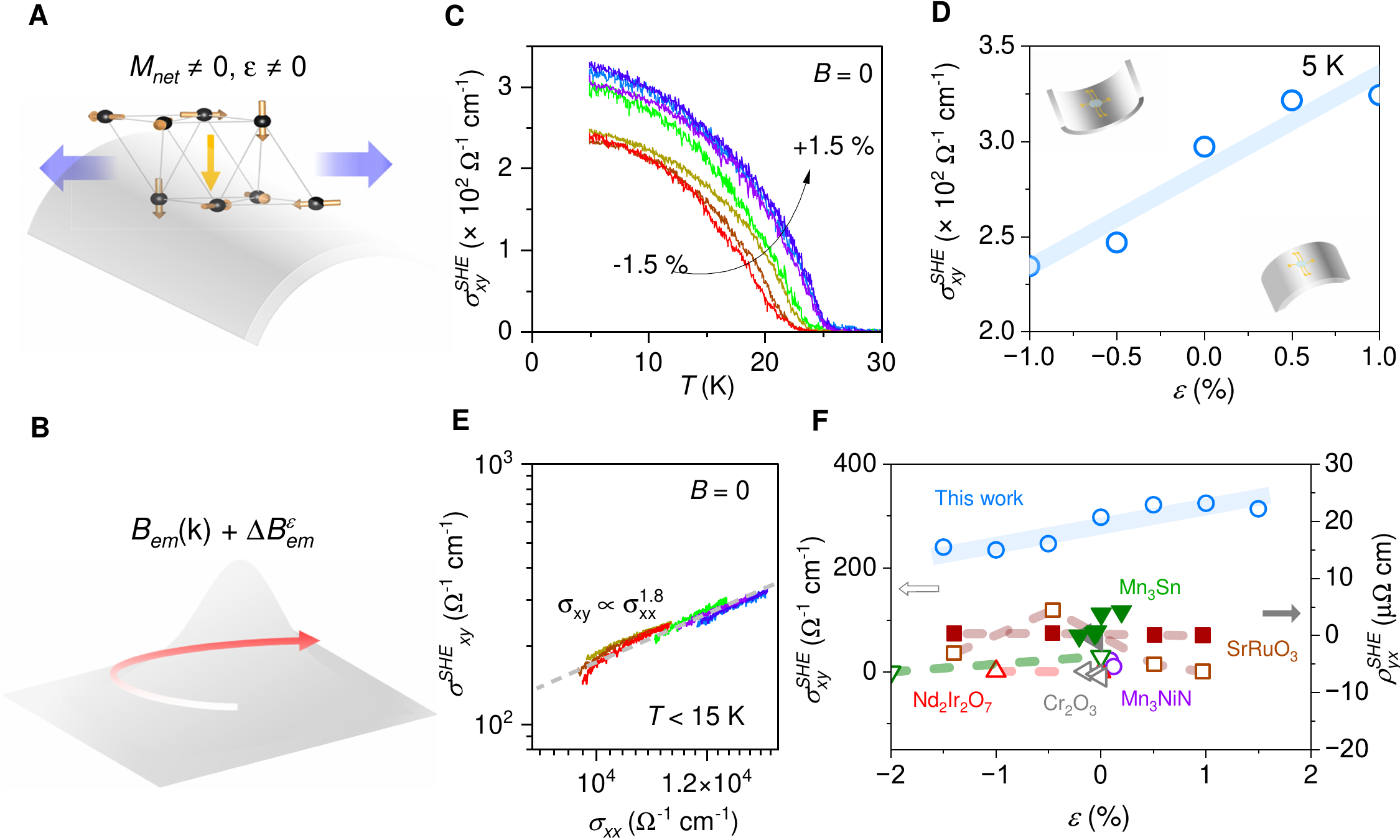}
\caption{{\bf Piezomagnetic control of spontaneous Hall effect.} \\
{\bf (A)} Generation of spontaneous magnetization ($M_{net}$) in real space by piezomagnetic effect. Orange arrow indicates a direction of $M_{net}$ under tensile strain. Under the strain, spins will change the canted angles, effectively creating net magnetic moment along the $z$ axis. {\bf (B)} Schematic of strain modulation of Berry curvature and spontaneous Hall effect. {\bf (C)} Temperature dependence of spontaneous Hall conductivity ($\sigma_{xy}^{SHE}$) recorded in various strain magnitudes (-0.015 $\leq$ $\epsilon$ $\leq$ +0.015). {\bf (D)} The strain variation of $\sigma_{xy}^{SHE}$ at the lowest temperature (5 K). {\bf (E)} $\sigma_{xy}^{SHE}$ - $\sigma_{xx}$ plot for all temperature below 15 K with various strain magnitudes (-0.015 $\leq$ $\epsilon$ $\leq$ +0.015). Dashed grey line refers to the exponent $\alpha$=1.8 (see text for details). {\bf (F)} Comparison of strain controlled spontaneous Hall conductivity (or resistivity) in various magnets with BTRS. The dashed lines correspond to data from multiple thin film samples with different strains, while the solid lines represent data obtained from variable strain on a single specimen. The open (filled) symbols indicate Hall conductivity (resistivity).}
\label{Fig3}
\end{center}
\end{figure}

It should be noted that the strain changes not only the exchange interactions in the effective spin model, which induces the magnetization and explains $T_N$ ($H_c$) modulations (real space picture, Fig.~\ref{Fig2}E and Fig.~\ref{Fig3}A), but also alters the electronic states and resultant electronic properties such as the magnetoresistance and Hall effect (including both normal and anomalous Hall coefficients) (momentum space picture, Fig.~\ref{Fig3}B, see also Fig. S4 in SI). Since the emergence of Berry curvature and associated spontaneous Hall effect is one of the prominent features in antiferromagnets with BTRS, we will focus on the strain modulation of these properties in this section. Such strain-tunable Berry curvature and anomalous Hall effect have been reported recently in several ferromagnets~\cite{Tian2021,Chi2023} and also in antiferromagnets with BTRS~\cite{Ikhlas2022,Boldrin2018,Guo2020}, but examples remain limited. Moreover, while many studies have been performed using epitaxial thin film samples with different substrate strains, in the present study we apply variable strain to a single device, allowing for the study of systematic strain effects. We note that recent strain-tuned Hall measurements on CoNb$_3$S$_6$ bulk crystal focused on a limited small strain window, where the response was attributed to an anomalous Hall effect~\cite{Chen2025}. In contrast, our measurements track the evolution of the Hall conductivity over a broader strain range, allowing us to establish robust scaling (see Fig.~\ref{Fig3}E and related discussion) behavior that constrains the underlying mechanism.

It is known that the $k$-space picture is useful to discuss and quantitatively estimate the $\sigma_{xy}$ in case that the size of the spin texture is comparable to crystallographic unit cell ($\lambda_s \sim a$)~\cite{Matsui2021,Khanh2025,Verma2022}. The spontaneous Hall conductivity in CoNb$_3$S$_6$ can be thus described as integration of momentum space Berry curvature over the occupied states in the momentum space; $i.e.$, $\int \frac{d^3 k}{(2\pi)^3}\,{B}_{em}({k})\,f_{\mathrm{FD}}(E_{{k}})$
 where $B_{em}$($k$) is the Berry curvature coming from the slightly gapped nodal planes which are the hot spots of emergent magnetic fields~\cite{Khanh2025}. Under the application of uniaxial strain, piezomagnetic effect can vary the solid angle of non-coplanar spin structure in real space and induce the finite magnetization, which can also contribute to the modulation of  Berry curvature in momentum space~\cite{Suzuki2017} ($B_{em}$($k$) + $\Delta B_{em}^\epsilon$) and $\sigma_{xy}$ ($B$=0) (see Figs.~\ref{Fig3} A and B). Figure ~\ref{Fig3}C shows the temperature variation of $\sigma_{xy}^{SHE}$ ($B$=0) under several strain values. It is found that $\sigma_{xy}^{SHE}$ ($B$=0) values are systematically modulated by strain as well as $T_N$ and $H_c$.  If we plot the magnitude of $\sigma_{xy}^{SHE}$ ($B$=0) at T = 5 K as a function of strain (Fig.~\ref{Fig3}D), it also follows the linear trend, representing a signature of piezomagnetic control of Berry-curvature-induced Hall conductivity. Several devices have been checked for the reproducibility of the results and directional dependence on the strain (see supplementary note 3). Note that the spontaneous Hall resistivity ($\rho_{yx}^{SHE}$ ($B$=0))  measured down to the lowest temperatures is found to be almost insensitive to the strain at lowest temperature and the change in the longitudinal resistivity $\rho_{xx}$ ($B$=0)) cause this modulation of $\sigma_{xy}$ ($B$=0)  (see Fig.S6 a, b). Alternative momentum-space descriptions of piezomagnetism have been proposed for itinerant antiferromagnets, particularly in systems with additional valley degrees of freedom~\cite{Ma2021,Hu2025}. In the present work, we follow the widely adopted spin-canting picture used for noncollinear metallic antiferromagnets (Mn$_3$Sn, etc.)~\cite{Ikhlas2022}, which provides a direct and experimentally relevant interpretation of the spontaneous Hall response. It is also noteworthy that the strain also modulates the ordinary Hall effect as well; compressive strain decreases the carrier concentration while tensile strain causes the opposite trend. In other words, the effective Fermi level is going down (up) by the compressive (tensile) strain, which may modify the anomalous Hall conductivity accordingly (see DFT results with different Fermi levels in Ref. \cite{Khanh2025}). However, our results show a behavior quite opposite to this scenario, highlighting that the modulation of Berry curvature, rather than a strain-induced shift of the effective Fermi level, is the main cause.
 
Figure 3E shows the $\sigma_{xy}^{SHE} - \sigma_{xx}$  plot, which is frequently used to discuss the mechanism contributing to anomalous Hall effect~\cite{Nagaosa2010}. The exponent $\alpha$ in relation $\sigma_{xy}^{SHE} \propto \sigma_{xx}^\alpha$ comes out 1.8, which suggests an intrinsic Berry curvature mechanism in bad metal regime~\cite{Capua2022}. This plot strongly suggests that the conductivity in the CoNb$_3$S$_6$ lies within the region of intrinsic mechanisms, and that the observed modulation of the spontaneous Hall effect is explained by Berry curvature. It is noteworthy that the scaling plots for all strained states collapse onto the unstrained curve, with neither the exponent nor the intercept exhibiting any systematic strain dependence. Since strain-induced magnetization would be expected to modify the amplitude of the spontaneous Hall conductivity through magnetization-related scattering (skew) or intrinsic contributions, the absence of any intercept ($\sigma_{xy}^{SHE}/M \propto \sigma_{xx}^\alpha$) or exponent modulation strongly disfavors a magnetization-dominated origin of spontaneous Hall conductivity ($\sigma_{xy}^{SHE}$)~\cite{Nagaosa2006,Bouma2020}. In Fig.~\ref{Fig3} F, we compare strain controlled $\sigma_{xy}^{SHE}$ (or $\rho_{yx}^{SHE}$) in various magnets with BTRS reported so far~\cite{Tian2021,Boldrin2019,Kim2020,Liu2023,Chi2023,Ikhlas2022}. It is evident that a very large strain modulation of physical properties has been achieved, reflecting both the nature of the van der Waals magnet and the characteristics of the strain-imprinting technique employed in this study.

\section{Effect of uniaxial strain on antiferromagnetic transport in \NoCaseChange{CoTa$_3$S$_6$}}
\begin{figure}
\begin{center}
\includegraphics*[width=14cm]{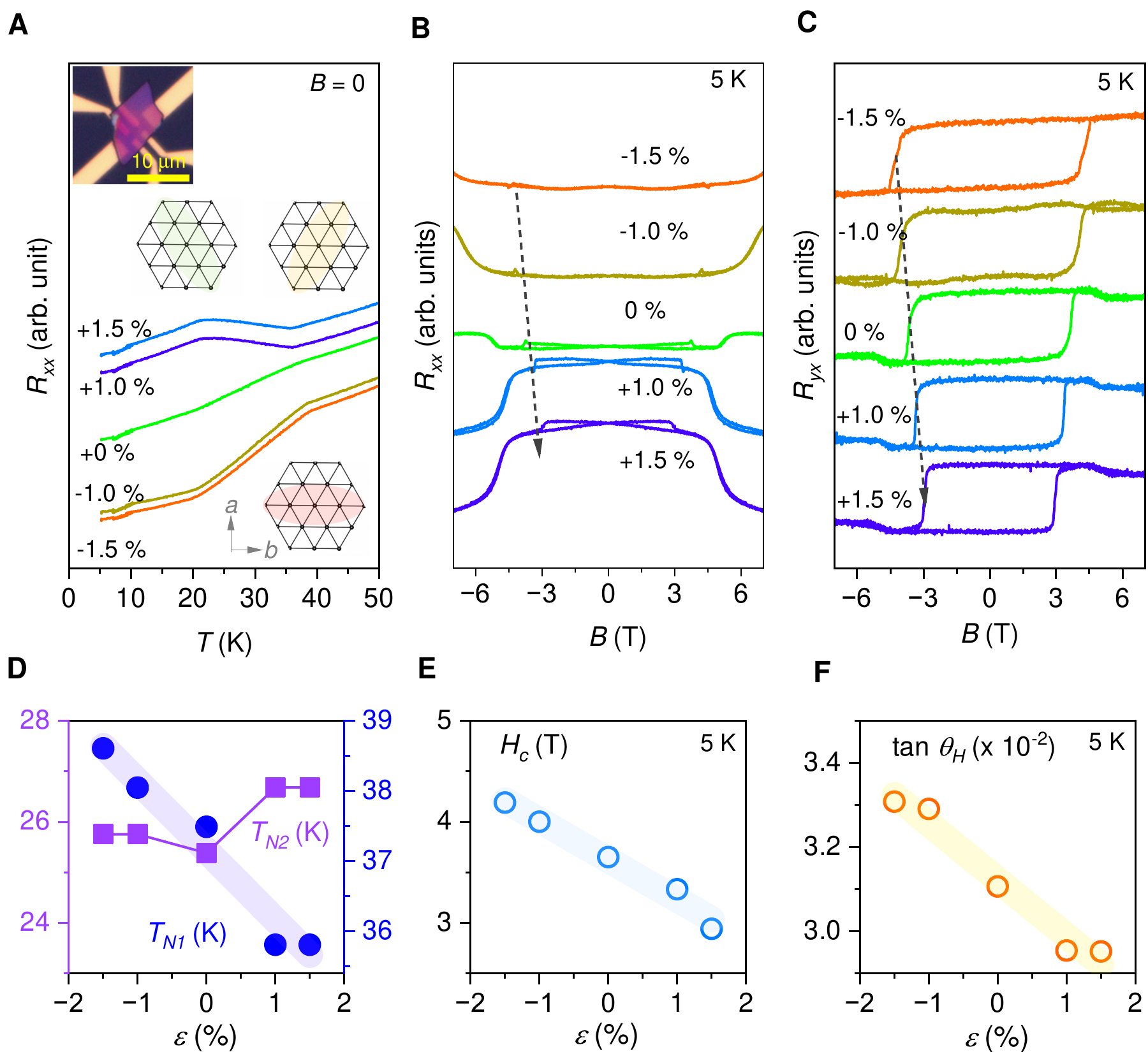}
\caption{{\bf Strain-tunable anisotropy and antiferromagnetic transport in CoTa$_3$S$_6$} \\
{\bf (A)} Temperature dependence of longitudinal resistance ($R_{xx}$) measured under various strain magnitudes under $B$ = 0. Insets show the optical microscope image of the device and schematics of nematic domains. {\bf (B)})-{\bf (C)} Magnetoresistance B and Hall effect C at 5 K recorded under various fixed strains. Dashed line showing the evolution of coercive field with strain. {\bf (D)}-{\bf (F)} Strain dependences of Neel temperatures ($T_{N1}$ and $T_{N2}$), coercive field ($H_c$), and Hall angle (tan~$\theta_H$). $H_c$ and tan~$\theta_H$ are extracted from the data at 5 K.}
\label{Fig4}
\end{center}
\end{figure}

We have further performed a similar set of measurements in CoTa$_3$S$_6$. It has crystal and magnetic structure similar to CoTa$_3$S$_6$ but shows notable differences; two types of Neel orderings have been observed at $T_{N1}$ $\sim$ 37 K and $T_{N2}$ $\sim$ 26 K~\cite{Takagi2023}.  At $T_{N1}$ it undergoes from paramagnetic phase to collinear antiferromagnetic phase. Within the temperature range between $T_{N1}$ and $T_{N2}$ ($T_{N2}$ $<$ $T$ $<$ $T_{N1}$), time reversal symmetry is preserved, and no spontaneous Hall effect is observed, $i.e.$, $\sigma_{xy}$ ($B$=0)=0. By further cooling down below $T_{N2}$, it enters antiferromagnetic states with BTRS which shows finite $\sigma_{xy}$ ($B$=0)~\cite{Takagi2023,Park2023}. Figure~\ref{Fig4}A shows the longitudinal resistance as a function of temperature under different uniaxial strain magnitudes. Interestingly, the trend of $R_{xx}$ (T)  in the region between $T_{N1}$ and $T_{N2}$, $i.e.$, whether $R_{xx}$ (T) decreases or increases with T, seems switchable with strain. It may be related to the recently suggested nematic phase~\cite{Feng2025} where in-plane uniaxial strain breaks the six-fold rotational symmetry of the lattice and thus can select nematic domains (insets of Fig.~\ref{Fig4}A).  Note that Ref.~\cite{Feng2025} mainly focused on resistivity anisotropy, and the influence of strain on the Hall conductivity, which is the central topic of this study, remained unexamined due to the limited strain values employed. It is also worth noting that the sign of the spontaneous Hall effect is not affected by this strain-switchable nematic phase (see Fig.~\ref{Fig4}B). Prominently, also in this CoTa$_3$S$_6$ sample, a linear modulation of $T_{N1}$, coercive field $H_c$, and Hall angle have been observed (Figs.~\ref{Fig4}D-F). We believe that it can be explained by similar piezomagnetic mechanism as discussed in CoNb$_3$S$_6$ samples. However, unexpectedly, the sign of the linear slope of the strain modulation in CoTa$_3$S$_6$ is opposite to that in CoNb$_3$S$_6$. We have checked the different strain direction and temperature dependence of the slope, but the results are similar (see supplementary material note 4 and Fig. S10). It indicates that the sign of the piezomagnetic tensor is opposite between CoNb$_3$S$_6$ and CoTa$_3$S$_6$, whose origin should be further pursued in the future.
	
\section{Conclusions} 
We demonstrate the piezomagnetic control of transport properties by using the van der Waals non-coplanar antiferromagnets CoNb$_3$S$_6$ and CoTa$_3$S$_6$ with BTRS. Antiferromagnetic transition temperature and coercive field are linearly modulated by uniaxial strain, which can be explained by strain tuning of exchange interactions and is consistent with piezomagnetic effect. Moreover, spontaneous Hall effect can be also tuned by strain, indicating the geometric electronic properties and associated transport can be also controllable via piezomagnetic effect. Present results indicate the potential for controlling the transport properties of antiferromagnetic materials through strain and pioneer emergent functionalities of van der Waals magnets.

\section{Materials and Methods}
\subsection{1.	Crystal growth and characterization}
Polycrystalline samples of CoNb$_3$S$_6$ and CoTa$_3$S$_6$ were synthesized from stoichiometric mixtures of elemental Co, Nb (Ta), and S powders sealed in evacuated silica ampoules. The mixtures were first heated at moderate temperatures (300$^o$ C for 2 days) and then at 900$^o$ C for several days, with intermittent regrinding to improve homogeneity and phase purity. Bulk single crystals were subsequently grown by chemical vapor transport (CVT) using iodine as the transport agent~\cite{Ghimire2018,Takagi2023}. The obtained single crystals adopt a thin-plate hexagonal morphology with the largest surface perpendicular to the crystallographic $c$-axis. The phase purity and structure were confirmed by powder X-ray diffraction, and crystal orientation was determined by Laue back-scattering. The compositions of Co, Nb/Ta, and S were verified by scanning electron microscopy (SEM, Hitachi S-4300) and energy-dispersive X-ray spectroscopy (EDX, Horiba EMAX x-act).

\subsection{2.	Device Fabrication}
Bottom electrodes devices were used for all the measurements. Flexible Polyethylene naphthalate (PEN) substrate of thickness 250 $\mu$m was used for the strain application. Substrates were cleaned in Iso-propyl alcohol (IPA) and coated with MMA/PMMA (methyl methacrylate/polymethyl methacrylate) layers with 3500 RPM as resist. Thin gold layer (10 nm) was deposited using ion-beam sputtering to dissipate the charge segregation during electron-beam lithography (EBL). A standard Hall bar pattern with six electrodes was designed using EBL (ELIONIC ELS-6600). After EBL, Au layer was etched using potassium Iodide solution and subsequently developed in cold IPA and water mixture. After development and UV-Ozone cleaning, Ti (3 nm)/Au(30nm) electrodes were sputtered followed by liftoff in Acetone. CoNb$_3$S$_6$ (CoTa$_3$S$_6$) flakes were mechanically exfoliated on the SiO$_2$/Si substrates in the glove box environment. Selected flakes were transferred on the pre-patterned PEN substrates using dry transfer method~\cite{Castellanos-Gomez2014} with polycarbonate (PC) (from Sigma-Aldrich) film at 180$^o$ C. Two types of devices with different crystal orientations (strain parallel to straight edge and perpendicular to straight edge) were prepared and measured in this study. Substrate was washed in Chloroform to dissolve the PC. Subsequently a thick layer (200 nm) of PMMA is deposited on the flake to protect it from oxidation and efficient strain transfer to the flakes~\cite{Gaikwad2022}. Thickness of all the flakes were measured using atomic force microscopy by employing Bruker Dimension Icon in Scan Asyst mode with soft X-ray bombarding to dissipate the charge build up on PEN.

\subsection{3.	Strain application using flexible substrate}
We have used the home-made setup to apply the uniaxial strain to all devices~\cite{Pu2021}. PEN substrate with the flake and the bottom electrodes were mounted between bending stage, of various radius of curvature, with the help of screws. The maximum applied strain was 2 $\%$, estimated using formula $t ⁄ 2R$, where $t$ is the substrate thickness (250 $\mu$ m) and R (6.25 mm) is the radius of the curvature of the bending stage.

\subsection{4.	Transport measurements}
The transport measurements of the devices were performed in a 9 Tesla Quantum Design, Physical Property Measurement System (PPMS) using a Keithley 2612 to source the current and two Agilent 34410 to record the transverse and longitudinal voltages simultaneously. We have recorded the resistivity ($\rho_{ij}$) as a function of temperature, from 5 K to 300 K using standard four-probe technique. The measurement of magnetoresistance (MR) ($\rho_{xx}$  ($B$)) and Hall ($\rho_{yx}$ ($B$)) have been performed at fixed temperatures. The recorded MR and Hall measurements were symmetrized and anti-symmetrized using relations $R_{xx}=(R_{xx} (+B)+R_{xx} (-B))/2$ and $R_{yx}=(R_{yx} (+B)-R_{yx} (-B))/2$ to eliminate the contact asymmetry, respectively. The Hall conductivity ($\sigma_{xy}$) were calculated using the following relation:

\begin{equation}
\sigma_{xy} = \frac{\rho_{yx}}{\rho_{xx}^2+\rho_{yx}^2}
\end{equation}

The spontaneous Hall resistivity as a function of temperature was measured by first cooling the sample under a magnetic field of B = 9 T and 7 T (if not mention specifically) and then recording the resistivity under zero magnetic field while warming process.

\section*{Data availability}

The data presented in the current study are available from the corresponding authors on reasonable request.

\section*{Author contributions}

A.A. and T.I. conceived the idea and designed the experiments. T.K. H.O. J.P. built up the strain setup. A.A. and M.T. prepared the devices and measured the experimental data. N.D.K., A.K. and S.S. synthesized the crystals. A.A., M.T., T.I., and S.S. wrote the manuscript with the input of all authors.

\section*{Acknowledgements} T.I. was supported by Japan Society for the Promotion of Science (JSPS) KAKENHI (Grant Numbers JP23H00088, JP24H01176, JP25H00839, JP25H02117), JST FOREST (Grant Number JPMJFR213A), and Murata Science and Education Foundation (Grant Number M24AN111). N.D.K was supported by JSPS KAKENHI Grant Numbers JP25K00956 and JP23K13069. J.P. was supported by JSPS KAKENHI (Grant Numbers 21H05232, 21H05236) and JST FOREST (Grant Number JPMJFR223Z). A.A. acknowledges support from JSPS KAKENHI (Grant Number 23KF0195).

\section*{Additional information}
Supplementary Information is available in the online version of the paper. Correspondence and requests for materials should be addressed to T.I. and A.A.

\section*{Competing financial interests}
The authors declare that they have no competing financial interests.

\clearpage

\end{document}